\documentclass{aa}
\usepackage[varg]{txfonts}
\usepackage{graphicx}
\usepackage{natbib}
\usepackage{float}
\begin{document}
	\title{The common trend of saltation particles on the surface of fast-rotating asteroids}
	\author{Zhijun Song\inst{1} \and Yang Yu\inst{1} \and Bin Cheng\inst{2} \and Jing Lv\inst{1} \and Hexi Baoyin\inst{2}}
	\institute{School of Aeronautic Science and Engineering, Beihang University, Beijing 100191, PR China \and School of Aerospace Engineering, Tsinghua University, Beijing 100084, PR China}
	\abstract{An asteroid spun up to its critical limit has unique surface mechanical properties that its gravity and the centrifugal force largely balance, creating a relaxation environment where low-energy events such as mass shedding may trigger subsequent long complex motion of an asteroid's regolith grains. Exploring such an evolution process may provide key clues for understanding the early formation of multi-asteroid systems.}
	{This paper investigates the complex evolution process of loose particles becoming triggered by shedding events and the dependency of their dynamical propagation on the contact mechanical properties of the asteroid surface.}
	{We present a numerical model for tracking the trajectory of a shed particle that considers the collision between the particle and the surface of an asteroid. Monte Carlo simulations are performed to reflect the statistical behavior of shed particles. We also introduce zero-velocity surfaces to our data-based analysis in order to reveal the intrinsic invariance of the evolutionary processes. We used the average mechanical energy of the particle cloud to check the connection between contact property and the temporal-spatial distribution of the shed particles.}
	{We sketch a common evolutionary path of the particle in the vicinity of a fast-rotating asteroid, that is, particles dislodged from the unstable region will eventually enter, through several collisions with the surface, non-return orbits that launch from the minimum geopotential area of the unstable region. The common trend is independent of any particular asteroid morphology, and all shed particles (no matter where they originate from) enter the same evolutionary path. We also find that the orbital energy of the particle cloud is statistically independent of the surface contact property, meaning that the collision coefficient of restitution is a nonsensitive parameter in the outward spreading process of the shed particles.}{}

	\maketitle
	
\section{Introduction}

Asteroids holding a "rubble-pile" structure have been studied for decades as an unprecedented state of mass aggregation \citep{Kevin_Rubble_pile,D.Hestroffer_Small}. A connection between the shapes of rubble-pile asteroids and their spin rates has been noticed \citep{Holsa_2001,Holsa_2004,Harris_Shape_spins,Hirabayashi_Spin-driven}. For example, top-shaped asteroids are commonly observed among the population of rapidly spinning asteroids \citep{Ostro_2006_science,Busch_2011,Naidu_Didymos_Shape,Watanbe_2019_Ryugu}. Several in situ space missions observed their target bodies as being covered with regolith grains of an extensive variety of sizes \citep{Regolith_science,Della_Properties_Bennu}. Yet mass transport or migration of the regolith layer has not been directly observed, and only circumstantial evidence presently supports such processes \citep{Bennu_eject_science,311p_activity}. Although these local migration events happen in a relatively short time span, they probably conceal important clues about the evolution of asteroids \citep{Cheng_B_nature}.

Regolith on non-cohesive surfaces have been shown to migrate from the mid-latitudes of an asteroid toward the equator region when the asteroid has a rapid spin rate \citep{DJ_Scheer_Landslide_and_Mass,Cheng_B_nature,Scanchez_Cohesive_regolith_2020,Hirabayashi_Spin-driven}. However, the equator is not always the final destination of these landslide failure materials. \citet{DJ_Scheer_Landslide_and_Mass} have noted that cohesionless regolith landslides at elevated spin rates may enter orbit once they migrate down to the equatorial region, as the net gravitational and centripetal acceleration at asteroid's equator is near zero. Differing from the ejecta of high-speed bombardments, the movements of regolith materials triggered by such landslides failures only gain low relative velocities \citep{YuY_2019,311p_activity,Kleyna_2019_Gault_Activity}. Therefore, the shed mass will become trapped in the potential well and be bound to the asteroid for a relatively long time. The subsequent activities of the shed materials essentially depend on the interactions between the lofted particles and the progenitor body, which are dominated by the perturbed gravitational field and the irregularly shaped surface of the asteroid. Given the complexity of this process, so far, how the shed particles will evolve in the vicinity of the asteroid is still an open question \citep{DJ_Scheer_Landslide_and_Mass}. Furthermore, this question is of particular interest for inferring the fission phase of a rubble-pile progenitor, and for the understanding of the initial formation stage of a multi-asteroid system. Thus, in this paper, the common evolutionary path for regolith particles shed from the equatorial region of fast-rotating asteroids is sketched through theoretical analysis and numerical simulation, in order to provide some insights into multi-asteroid system formation.
		
Efforts have been devoted to understanding such processes over the past decade. In particular, \citet{Walsh_Rotational_breakup} have demonstrated that it is possible for shed mass from the equator of a critically spinning asteroid to aggregate into a satellite when the material is collisionally dissipative. While this result provides a mechanism for the formation of a binary system, it does not capture details of the shed mass moving around in this system. Furthermore, a rubble-pile asteroid structure implies that there is a specific spin rate at which some component of the asteroid will go into orbit around the rest of the material, such as when the centripetal acceleration exceeds gravitational attraction\citep{Scheeres_Minimum_2009}. In this context, the shedding conditions for cohesionless regolith on fast-rotating asteroids have been derived for both irregularly and simplistically shaped models \citep{Hirabayashi_Analysis_2014,DJ_Scheer_Landslide_and_Mass,Hirabayashi_Failure_2015,YuY_2018,YuY_2019,Scanchez_Cohesive_regolith_2020}. \citet{Tardivel_Equatorial_cavities_2018} reported a possible formation mechanism for the equatorial cavities of asteroids 2008 EV5 and 2000 DP107 and proposed a plausible mechanism for binary formation based on the hypothesis that cavities could be the result of rotational fission. All of these studies have focused on how regolith particles could shed from the surface, but they do not discuss the dynamical fate of the shed materials of asteroids in detail. 

There have also been works carried out to investigate the fate of the low-speed ejecta from the surface of asteroids. \citet{McMahon_Particles_Bennu_2020} has discussed the fate of the low-speed ejecta from an asteroid surface under conditions similar to those observed at Bennu and explored how the parameters of the dynamical system and the initial particle conditions influence particles moving around Bennu. \citet{Nakano_MassSheding} proposed their interpretation of the process that produced the Geminid meteor stream by investigating the mass shedding from the asteroid (3200) Phaethon. We note that these studies assume the particles directly stick to the asteroid once they re-collide with its surface and have not considered the continuous interplay between the lofted particles and the asteroid nor connection such interplay with the dynamical fates of the lofted mass. For example, the suborbital trajectories of the shed particles could repeatedly undergo small hops (a process of shedding, orbital movement, and collision with the surface) and bouncing (i.e., re-entering orbit), as they only gain low relative speeds. Therefore, if the particles are assumed to stick to the surface of the asteroid after collision, it would be impossible to capture such a pattern of motion and that may cause the distribution of shed particles near the asteroid to deviate from the actual situation. We also note that although the mechanism is different, there is an intrinsic similarity between the repeat hop motions of shed particles on the surface of an asteroid and both the "tidal saltation" process\citep{Harris_Shape_spins} and the saltation process described by \citet{owen_1964_saltation}. Therefore, in this study, we use the word "saltation" to describe the post-shed motion of regolith particles. 
	
The purpose of this study is to understand the complex saltation process of the shed materials from the unstable regions of an asteroid and the dependencies of their dynamical propagation. Without loss of generality, the shape and physical parameters of the primary of the Didymos system were chosen to serve as a hypothetical example for numerical validation. The Didymos has a rotation period close to its critical spin limit and is thus likely to shed mass from its equatorial bulge. We took into account the collision of the particle with the asteroid surface, the asymmetric surface morphology, and the nonspherical gravitational field of the target asteroid. Unstable regions of the surface were first profiled to determine plausible shedding locations. Monte Carlo simulations were implemented to assess the temporal-spatial evolution of the shed particles as a function of the gravitational field, the surface morphology, and the collisional properties of the asteroid. The remainder of the paper is organized as follows. In Sec. \ref{S_2}, we briefly introduce methods and theories involved in this study, including the identification of the unstable region, the governing equation of the particle motion, and the numerical simulation scheme. Sec. \ref{S_3} presents the results of the numerical simulations designed to track the saltation routes of particles and a discussion on their collisional dependency. Finally, in Sec. \ref{S_4}, we provide concluding remarks. 
	
\section{Methodology} \label{S_2}

The target asteroid is assumed to be a homogeneous rigid body, that uniformly rotates about its maximum inertial principal axis. The shed regolith materials are considered to be loose cohesionless particles whose ballistic motion is dominated by the gravitational force of the asteroid. In other words, no extra perturbations were considered. The trajectories of particles are represented in the body-fixed frame of the asteroid, with the origin located at the mass center and the x, y and z axes consistent with the minor, intermediate, and maximum principal axes, respectively. We also assumed collisions between the shed particle and the asteroid surface are an instantaneous process with energy loss determined by the coefficient of restitution.

\subsection{Identification of unstable regions}

Unstable regions of the asteroid surface were first identified in order to create a plausible shedding scenario and determine the post-collision behaviors of the particles. In this study, we adopt the definition of the geopotential introduced by \citet{Scheeres_1998_Orbits} and define the local regions where the gradient direction of the geopotential is outward as the unstable region. Thus, the cohesionless regolith grains initially resting in such regions will be shed from the surface due to grains feeling a centrifugal acceleration that exceeds the gravitation attraction \citep{YuY_2018}. The topographic condition of local unstable regions can be formulated as
    	\begin{equation}\label{1}
    	 	l=\nabla{V_\mathrm{s}(\vec{r})}\cdot \vec{n} 
    	,\end{equation}    
where $\vec{r}$ indicates the position vector of particles in the fixed-body frame, $\vec{n}$ is the unit normal vector of the local surface (pointing outward), $\nabla$ represents the gradient operator, and $V_s\vec(r)$ indicates the geopotential on the asteroid's surface with position vector $\vec{r}$. The geopotential is defined as		
	\begin{equation} \label{2}
		V(\vec{r})=-\frac{1}{2}(\vec{\omega} \times \vec{r}) \cdot (\vec{\omega} \times \vec{r})+U(\vec{r})
	,\end{equation}	
where $\vec{\omega}$ is the angular velocity vector, which is a constant in the context, and $U(\vec{r})$ indicates the potential of the gravity of the point with position vector $\vec{r}$, which is computed by the polyhedral method \citep{Werner_1996}. Therefore, regions subject to $l\leqslant0$ are unstable regions in the sense we mentioned previously. That is, they are regions where the particles would experience outward acceleration, separating the particles from the body of the asteroid. 

\begin{table}[htb]  
	\centering
	\caption{Physical parameters of Didymos}   
	\label{table:1}                                              
	\begin{tabular}{c c c c}                                
		\hline\hline                                          
		Mass              &   Period         & Mean radius      & Bulk density \\
		$\mathrm{m}$ $\mathrm{[kg]}$  & $\mathrm{T}$ $\mathrm{[h]}$  & $\mathrm{R}$ $\mathrm{[m]}$  &$\mathrm{\rho}$ $\mathrm{[kg}$ $\mathrm{m^{-3}]}$\\
		\hline    
		$5.384^{11}$     & $2.26$           & $387.5$          & $2146$     \\
		\hline
	\end{tabular}
\end{table} 

\begin{figure}[htb]
	\resizebox{\hsize}{!}{\includegraphics{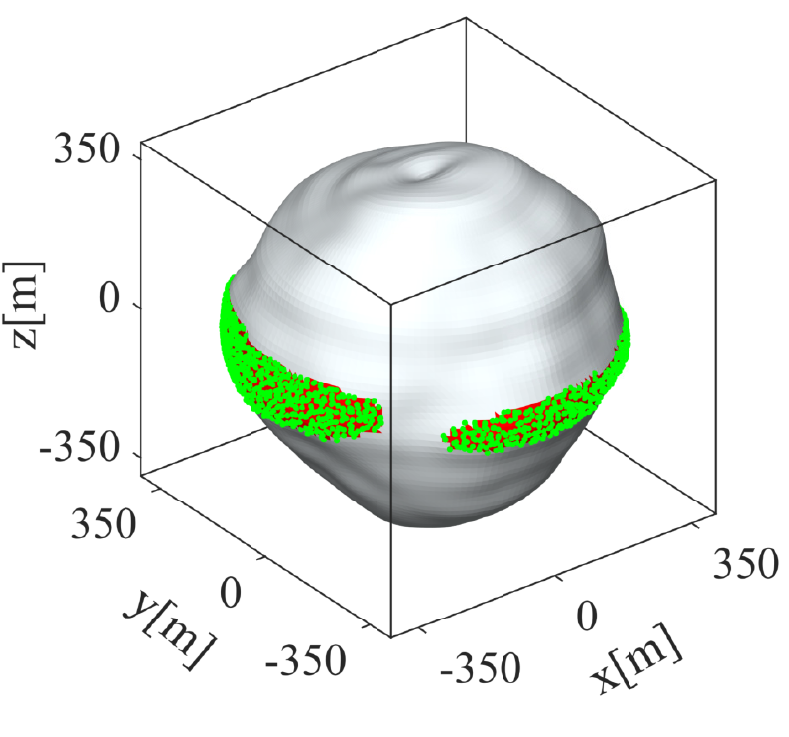}}  %
	\caption{Shape of Didymos and the initial distribution of particles on its surface. The green scattered points indicate the initial positions of particles, and the red region corresponds to the unstable region.}
	\label{fig:1}
\end{figure}  

\begin{figure}[htb] %
	\resizebox{\hsize}{!}{\includegraphics{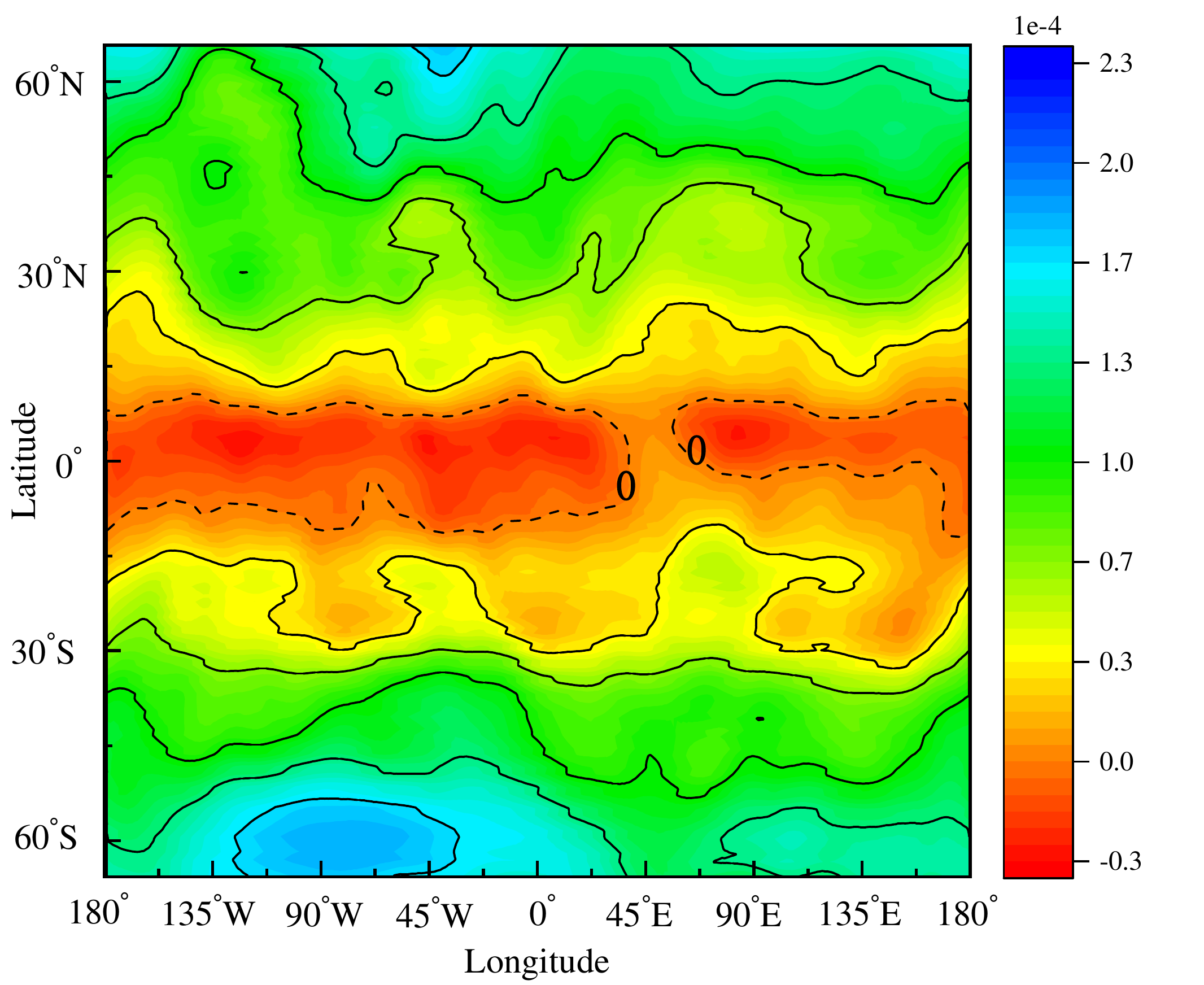}}  %
	\caption{Solutions of the unstable regions of Didymos. The full color spectrum corresponds to the minimum to maximum values of $l$ as defined in Eq.\,(\ref{1}). The red color represents the unstable region $l\leqslant0$, and the blue color indicates the stable regions $l\geqslant0$. The solid curves indicate the contour lines of the value of $l$, and the dashed curves indicate that the value of $l = 0$, which corresponds to balanced, normal components of the gravitational and centrifugal forces.}
	\label{fig:2}
\end{figure}
    
\subsection{Equations of motion}

Based on the mechanical environment described in the previous section, the equation of orbital motion of a particle is represented in the body-fixed frame as 
    \begin{equation}\label{3}
    	\ddot{\vec{r}}+2\vec{\omega}\times\dot{\vec{r}}+\vec{\alpha}\times\vec{r}+\vec{\omega}\times(\vec{\omega}\times\vec{r})=-\nabla U(\vec{r})
   ,\end{equation}    
where $\dot{\vec{r}}$ and $\ddot{\vec{r}}$ respectively indicate the velocity and acceleration of the particle with regard to the body-fixed frame, and $\vec{\alpha}$ is the angular acceleration vector. Since we assume Didymos is uniformly rotating, $\vec{\alpha} = 0$ in this paper. Then by substituting Eq.\,(\ref{2}) into Eq.\,(\ref{3}), Eq.\,(\ref{1}) becomes   
    \begin{equation}\label{4}
    	\ddot{\vec{r}}+2\vec{\omega}\times\dot{\vec{r}}=-\nabla V(\vec{r})
   .\end{equation}
    
There is an integral of generalized energy in this dynamical system \citep{YuY_Orbital_dynamics}, which is expressed as     
    \begin{equation}\label{5}
    	J=\frac{1}{2}\dot{\vec{r}}\cdot\dot{\vec{r}}+V(\vec{r})
   .\end{equation}

The ballistic motion defined by Eq.\,(\ref{4}) maintains a constant Jacobi integral (generalized energy) in the case of no collision with the surface, and the constant value is also called the Jacobi integral. In the body-fixed frame, Eq.\,(\ref{5}) actually divides the exterior space into two parts: the allowable region, defined by $V(\vec{r})\leqslant J$, and the forbidden region, defined by $V(\vec{r})>J$ \citep{Scheeres_1996}. The two-dimensional surface defined by $V(\vec{r})=J$ in the three-dimensional space is known as the zero-velocity surface, which changes as the value of the Jacobi integral changes.
    
\subsection{Collision process}%

Collisions between the shed particle and the asteroid surface seem frequent in the current scenario, due to the low launching speed and the large unstable surface around the equator. Modeling the collisional process is therefore necessary to track the full trajectory of a particle. We adopted a simplified model to determine the velocity after a collision. In the model, the contact deformation is neglected and approximated to an instantaneous process. The velocity with respect to the surface of the asteroid after a collision can be solved using the classic collision theory of rigid bodies, and therefore be written as   
    \begin{equation}\label{6}
    	\left\{
    	\begin{aligned}
    		v^{+}_\mathrm{n}&=-\varepsilon_\mathrm{n}v^{-}_\mathrm{n}\\
    		v^{+}_\mathrm{\tau}&=\varepsilon_\mathrm{\tau}v^{-}_\mathrm{\tau}
        \end{aligned}
        \right.
   ,\end{equation}
where $v^{-}_\mathrm{n}$ and $v^{-}_\mathrm{\tau}$ are, respectively, the normal velocity and the tangential velocity regard to the surface of the asteroid before impact and $\varepsilon_\mathrm{n}$ and $\varepsilon_\mathrm{\tau}$ indicate the coefficient of restitution in the normal velocity direction and tangential velocity direction, respectively. Considering imperfect inelastic collisions, the coefficients of restitution are between zero and one, and the energy loss can be calculated accordingly.

\subsection{Simulation setup}

 \begin{figure}[htb]     	
	\resizebox{\hsize}{!}{\includegraphics{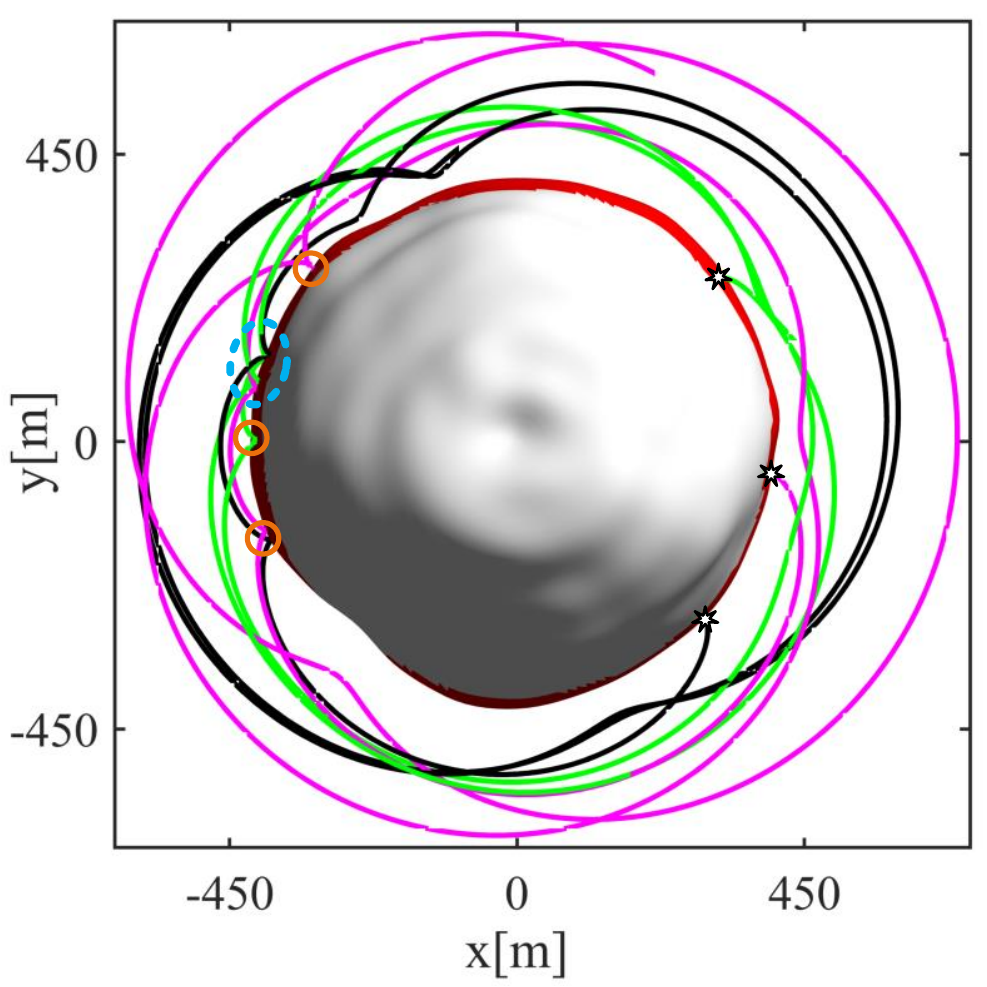}}  %
	\caption{Saltation routes of three different shed particles. The solid lines of different colors respectively correspond to three different saltation routes whose launching positions are denoted by star-shaped marks. The orange circles indicate collision sites on the different saltation routes. The area surrounded by the blue dashed line denotes the unified destination of the three saltation routes.}
	\label{fig:3}
\end{figure} 

We chose the primary of $65803$ Didymos as an example progenitor in our simulations because it has a typical top-like shape and a relatively high spin rate. We note, however, that we only take the Didymos primary as a hypothetical example of a single asteroid and that we dismiss its satellite, Dimorphos. We employed the physical parameters listed in Table\,\ref{table:1} and the shape model exported by \citet{Naidu_Didymos_Shape} (Fig.\,\ref{fig:1}).

We implemented a hypothetical shedding simulation following the previously described setups. A full-cycle procedure includes the following steps. First, the unstable regions are located using Eq.\,(\ref{1}). Given the observed values listed in Table\,\ref{table:1}, Fig.\,\ref{fig:2} illustrates the unstable region as a narrow band near the equator. A stable region was noted around $(55^{\circ}\mathrm{E},0^{\circ})$, which is a depressed area with low elevation. We observed that, as a general trend and due to the decrease of the centrifugal force, as the latitude increases, the surface stability of Didymos increases. This result is consistent with the previous investigation implemented by \citet{YuY_2018}. We noted that their research also revealed another unstable scene in a local surface where local slope failures accelerate particles to escape speed. However, such failures do not have to occur because of the complexity of local small-scale topographies. For example, regolith particles may be geometrically struck in the flow direction because of the obstruction of the local bulges or surface boulders. Nonetheless, consideration of such a scene is beyond the scope of this paper.
   
Second, Monte Carlo simulations are performed based on the identified unstable region. Tracing particles were initially located randomly across the unstable region, as shown in Fig.\,\ref{fig:1}. The particles are then released with zero velocity with respect to the local surface. Because of the outward net acceleration in the body-fixed frame, these particles shed and enter orbits in the vicinity of Didymos; and due to the low orbital energy, the particles frequently return to and collide with the surface, generating saltation trajectories near the asteroid surface. 

Next, month-long simulations are performed to find the dynamical fates of the lofted particles. In this paper, we determined whether a particle sticks to the surface after collision by its collision site and the magnitude of the velocity after collision. If the location of impact is in the stable region and the magnitude of the relative velocity after the collision is less than $0.001$ m/s, the particle is considered to have become stuck on the surface. In this paper, we simply assumed for the convenience of the numerical simulation, that when the rebounding height of a particle after collision with the surface is less than 0.1 times the particle radius, it may become stuck on the asteroid surface. Thus, the magnitude of the speed that may stuck on the asteroid surface after collision could be calculated by the following formula
 \begin{equation}\label{71}
	\vert v^{+} \vert=\sqrt{\mathrm{2gh}}
,\end{equation}
where $\mathrm{g}$ is the magnitude of the surface gravity acceleration of the asteroid and $\mathrm{h}=0.1\mathrm{R}$. For our simulation, the $\mathrm{g}\approx10^{-4}\mathrm{m\,s^{-2}}$ and  $\mathrm{R}=0.05\,\mathrm{m}$.

We considered eight groups of sample particles to be representative of regolith grains in different ranges of the coefficient of restitution. Each sample group includes 1,942 spherical particles with $5$ cm in radius. The coefficient of restitution in the normal velocity direction and tangential velocity direction for the particle was assumed to be equivalent and range from $0.15$ to $0.85$ with a span of $0.1$ to investigate the influence of collisional properties on the evolution of the shed particles.

\begin{figure}[htb] %
	\resizebox{\hsize}{!}{\includegraphics{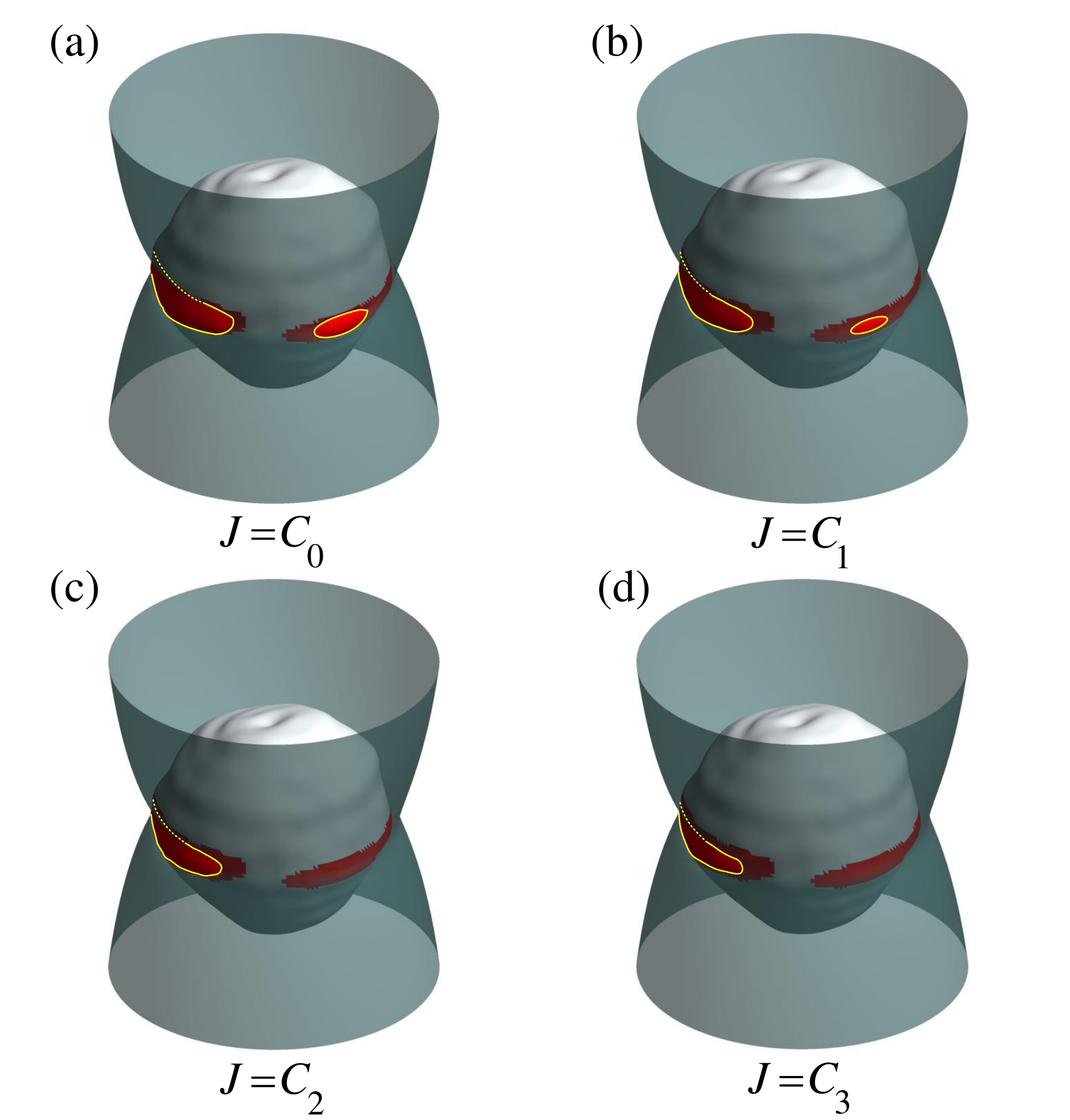}}  %
	\caption{Evolution of allowable regions on the surface of the asteroid caused by three collisions. The value of $\mathrm{C_0}=-0.1377\,\mathrm{J\,kg^{-1}}$ corresponds to the Jacobi integral of the particle at the initial moment, whereas $\mathrm{C_1}=-0.1381 \,\mathrm{J\,kg^{-1}}$, $\mathrm{C_2}=-0.1384\, \mathrm{J\,kg^{-1}}$, and $\mathrm{C_3}=-0.1387 \,\mathrm{J\,kg^{-1}}$ correspond to its Jacobi integral after three collisions, respectively. The semi-transparent gray surface represents the zero-velocity surface, and the yellow line represents the intersection of the zero-velocity surface and the asteroid surface. The red region corresponds to the unstable region. Regions protruding out of the zero-velocity surface and encompassed by yellow lines indicate the allowable regions on the surface of the asteroid.
	}
	\label{fig:4}
\end{figure} 

\section{Results} \label{S_3}

This section describes and analyzes the results of the numerical simulations. In Sec.\,\ref{3.1}, we characterize the saltation routes of the shed particle, and in Sec.\,\ref{3.2}, we discuss the parameter dependency of the particle dynamical propagation.  
    
\subsection{Saltation route analysis}\label{3.1} 

\begin{figure*}[h] %
	\centering
	\includegraphics[width=17cm]{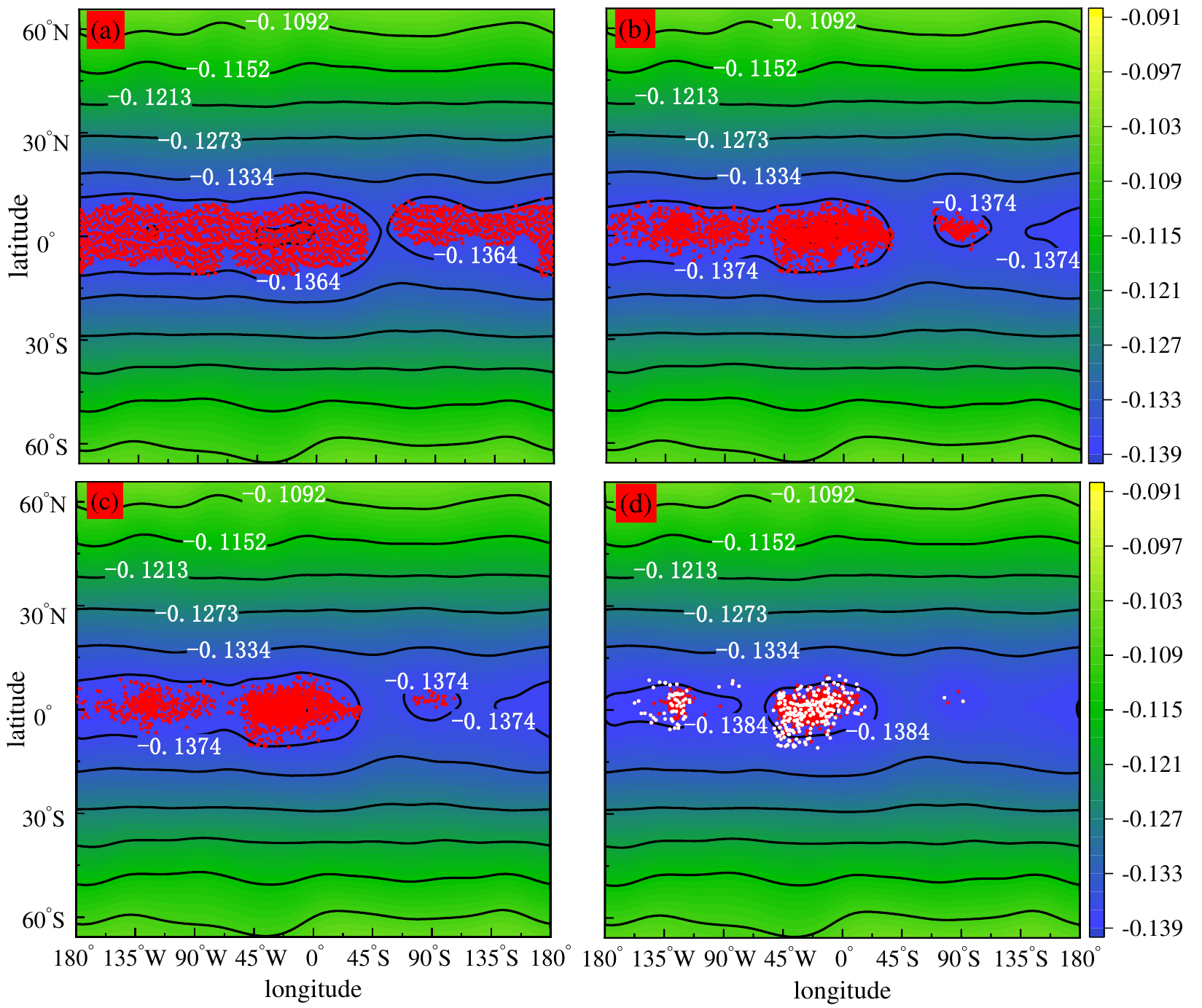}    	
	\caption{Evolution of collision sites with increasing number of collisions and the contour plot of the geopotential. The red points denote collision sites on the saltation routes, and the white points indicate the initial locations where particles directly enter orbit and no longer collide with the surface during the simulation. The black solid lines indicate the contour lines of the geopotential. The color spectrum (blue to yellow) corresponds to the geopotential ranging from $-0.1398$ to $-0.091$ $ \,\mathrm{J\,kg^{-1}}$. Panels a, b, c, and d show the location distribution of particles at the initial moment, the first collision, the second collision, and the last collision, respectively.  
	}
	\label{fig:5}
\end{figure*}

\begin{figure}[h]
	\resizebox{\hsize}{!}{\includegraphics{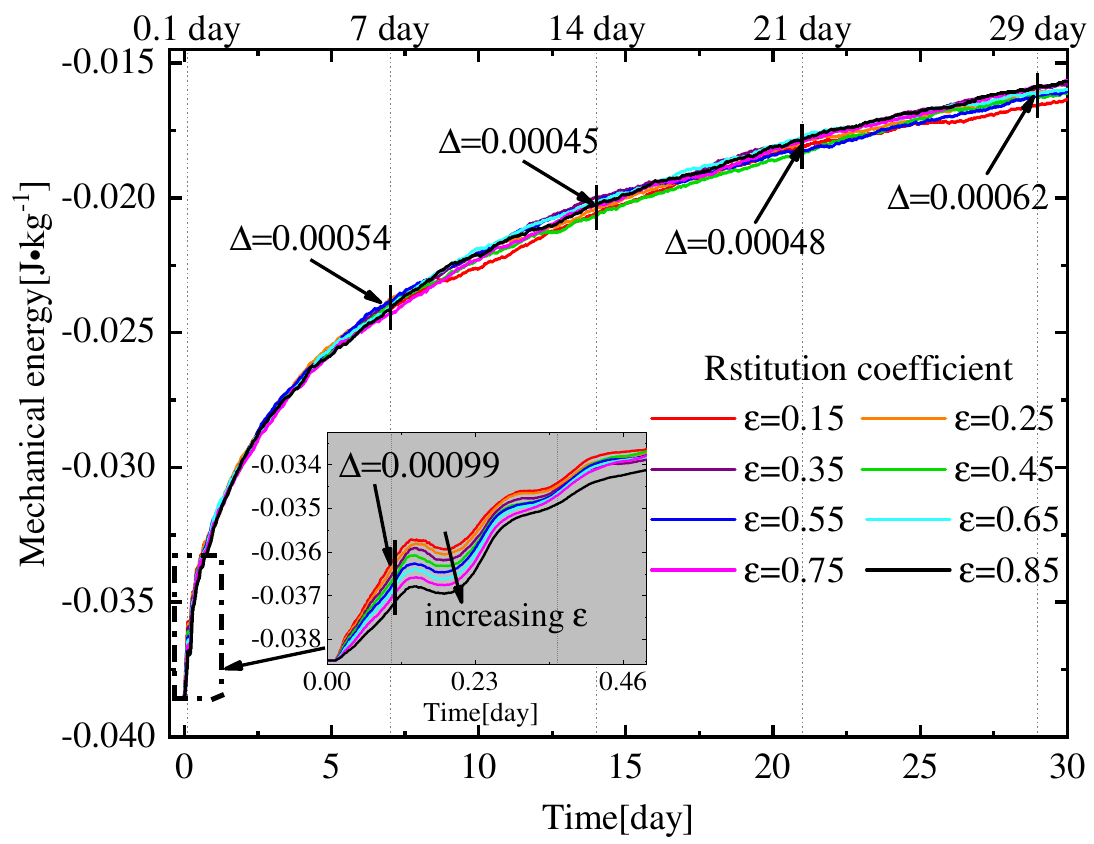}} 
	\caption{Evolution of the average mechanical energy of the particle cloud versus time within 30 days for eight groups of sample particles. The arrows indicate the difference of the mechanical energy at five different moments. The various curve colors represent the results of different coefficients of restitution and are coded as shown in the legend.}
	\label{fig:6}
\end{figure}

\begin{figure*}[h] %
	\centering
	\includegraphics[width=17cm]{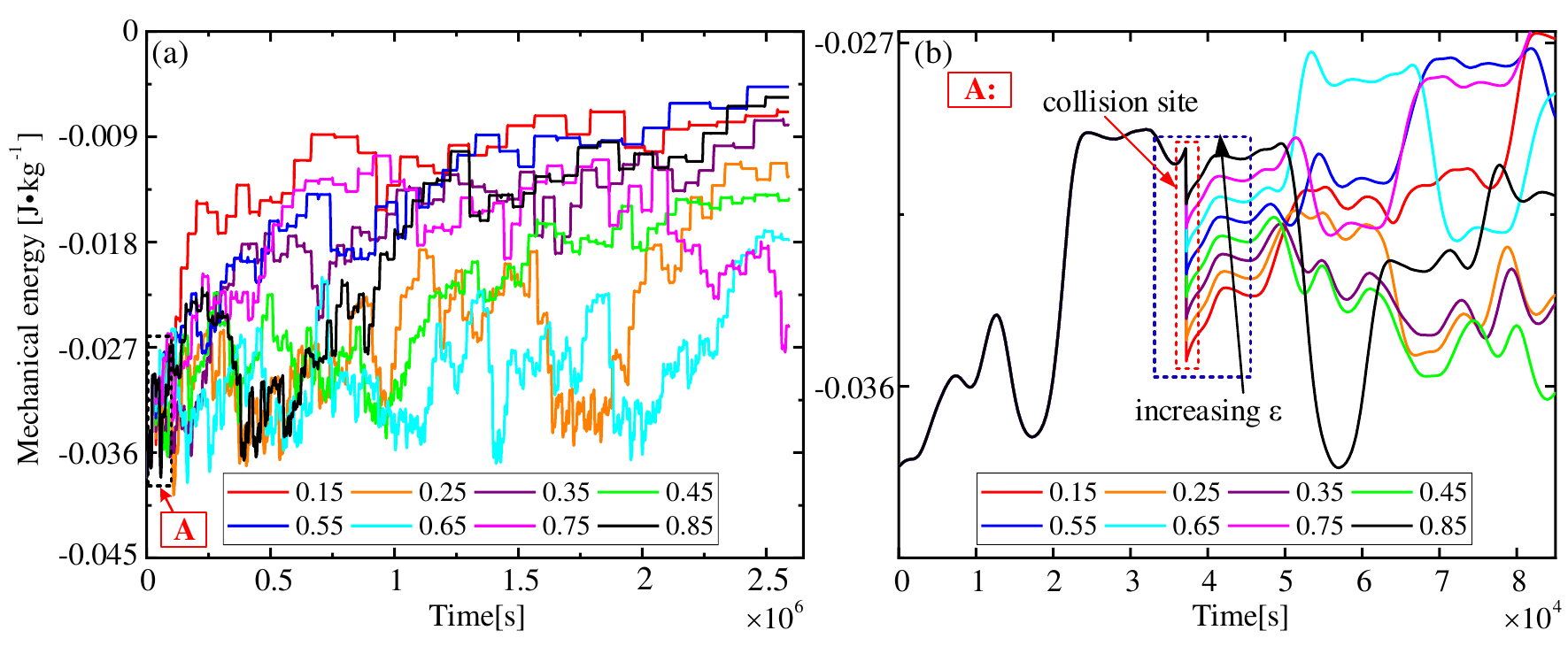}    	
	\caption{Change in the mechanical energy of particles with different coefficients of restitution versus time. Panels (a) and (b) correspond to the global change and the local change, respectively. The local change refers to the enlarged view of the part surrounded by the black dash-lined box in panel\,(a). The red dash-lined box marks the collision site, and the blue dash-lined box marks the range of time that coefficients of restitution keep monotonicity versus mechanical energy. The mechanical energy for different $\varepsilon$ is presented with different colors as indicated by the legend.}
	\label{fig:7}
\end{figure*}

\begin{figure}[h] 
	\resizebox{\hsize}{!}{\includegraphics{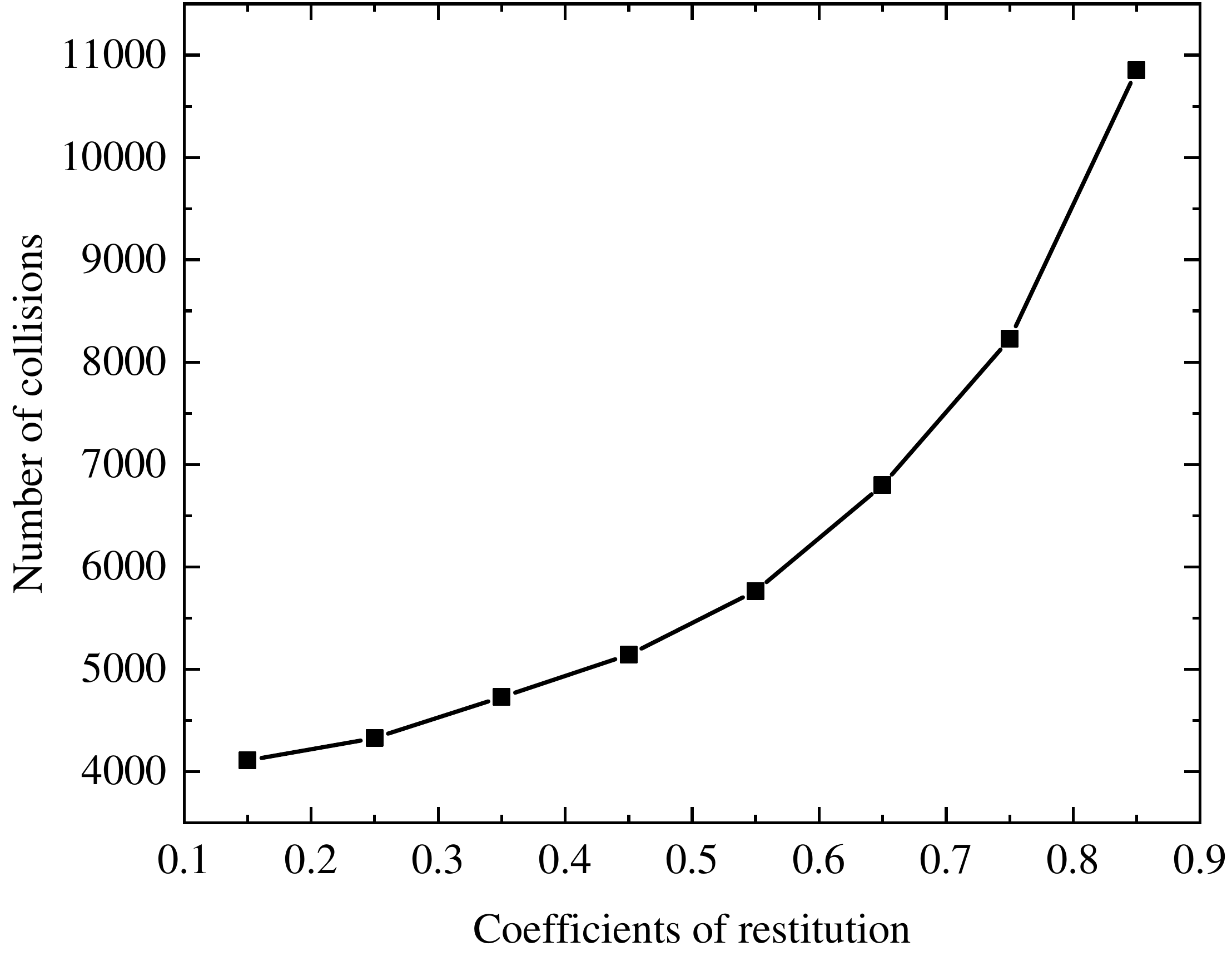}}  
	\caption{Number of particle cloud re-impact versus the coefficient of restitution.}
	\label{fig:8}
\end{figure}

\begin{figure*}[h]
	\centering
	\includegraphics[width=17cm]{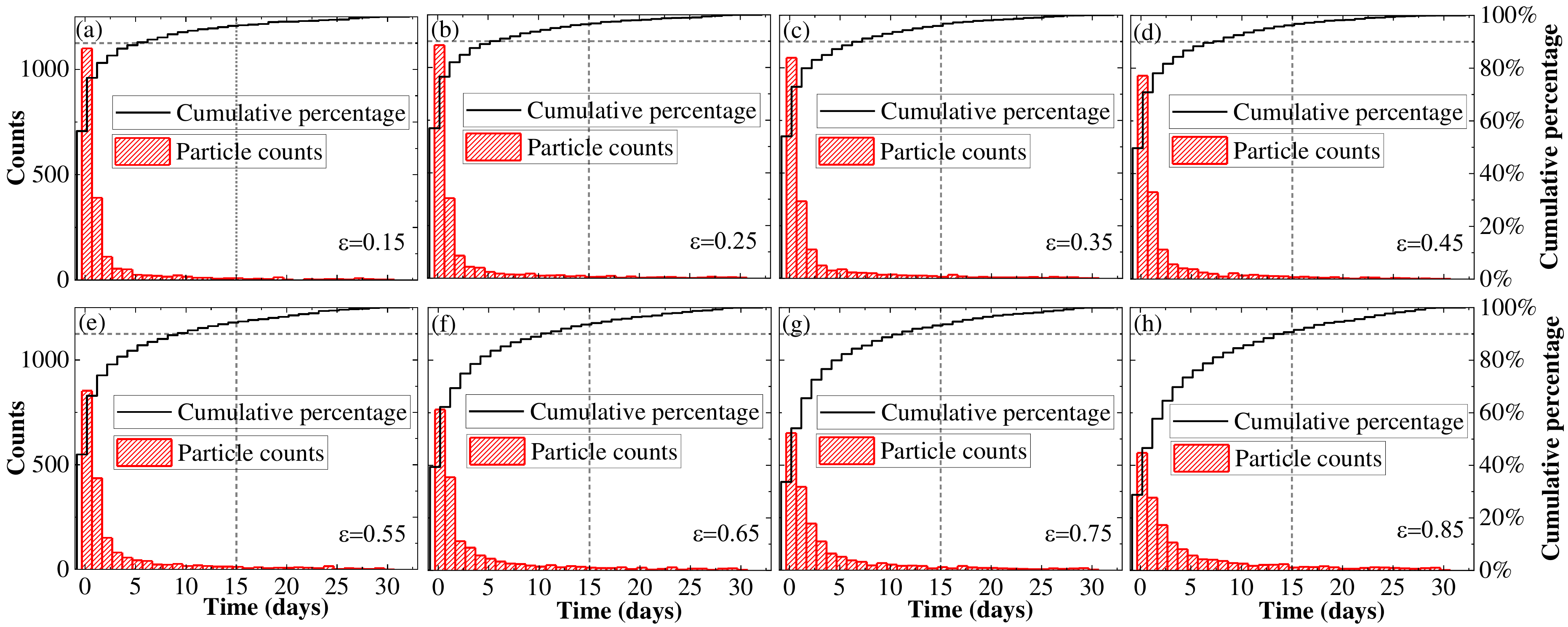}    	
	\caption{Time statistics of the particle with different coefficients of restitution entering the non-return orbits. The histograms show the distribution of the counts of particles entering the non-return orbits over the simulated time. The broken lines indicate the cumulative percentage of these particles. The green perpendicular dashed lines mark the moment of 15 days, and the horizontal dashed lines mark the 90\% cumulative percentage.}
	\label{fig:9}
\end{figure*}

We observed a common tendency among all the sampled trajectories: The saltation trajectories of regolith particles dislodged from the unstable region will eventually tend, through several collisions with the surface, to be near the minimum geopotential area of the unstable region. It is an abnormal phenomenon because all particles were randomly sampled across the unstable region, and the complex topography and nonspherical gravity should largely disturb the evolutionary trajectories. We took a detailed look at the energetic level and examined the motion of sampled particles in terms of Jacobi integral and zero-velocity surface \citep{Scheeres_1996}. Combining Eqs.\,(\ref{5}) and (\ref{6}), the variation of the Jacobi integral before and after the collision can be written as

    \begin{equation}\label{8}
    	\Delta{J}=\frac{1}{2}(\varepsilon^2-1)\dot{\vec{r^-}}\cdot\dot{\vec{r^-}}
    .\end{equation}

Since we assume that the shed particle has an imperfect inelastic, that is, the coefficient of restitution $\varepsilon$ is always less than one, $\Delta{J}$ must be negative according to Eq.\,(\ref{8}). And we note the fact that  $J$ remains invariant except in the collisional process. Therefore, during a saltation process composed of a consequence of collisions, the Jacobi integral $J$ will show a monotonous decrease as the number of collisions increases. This means the allowable region determined by the Jacobi constant will only change in the direction of a dropping $J$. Taking a general particle as an example, a qualitative description of the saltation route will be as follows: As the particle first detaches from the asteroid surface, the relative velocity is zero. Thus, the initial Jacobi integral is determined by the local geopotential $J_0 = V(\vec{r_0})$, where $\vec{r}_0$ is the launching position vector right at the intersection between the zero-velocity surface and the asteroid surface. The asteroid surface exposed to the allowable region $V(\vec{r})\leqslant J_0$ is accessible when the particle first returns, and the surface in the forbidden region $V(\vec{r})\geqslant  J_0$ will never be visited. After the first collision, the Jacobi integral decreases to $J_1<J_0$, causing a shift of the intersection between the zero-velocity surface and the asteroid surface. The asteroid surface exposed to the allowable region shrinks, as it is subject to $V(\vec{r})\leqslant J_1$. Thus, the next collision site of the particle $\vec{r_1}$, if it exists, will be located in a smaller surface region (i.e., $V(\vec{r_1})\leqslant J_1 < J_0 =  V(\vec{r_0})$). These processes repeat during the saltation process, and each collision leads to a shrinking of the accessible region from high to low geopotential until the overall asteroid surface drops into the forbidden region. This means that the saltation process ends, and the particle motion pattern switches from saltation on the surface to a long-term orbit motion. A series of collisions leads the accessible region shrink and provides a common mechanism governing the saltation routes. This mechanism drives the shed particles toward a common tendency formulated as ($\vec{r}_i$ indicates the $i^{\textup{th}}$collision site):

    \begin{equation}\label{81}
	V(\vec{r}_{n}) <V(\vec{r}_{n-1}) < \cdots < V(\vec{r}_1) < V(\vec{r}_0)
    .\end{equation}

Figure\,\ref{fig:3} gives a quantitative example of the saltation process in the top view. In the figure, three saltation routes are sketched from different initial positions with a coefficient of restitution of $0.45$. The routes exhibit the previously described trend: Collision sites of different trajectories migrate from high geopotential to low geopotential and converge around the minimum geopotential region after several collisions. We note that the particles were released with zero relative velocity and their motions were dominated by the gravitational force of the asteroid. Thus, the particles had a zero radial velocity with respect to the body-fixed frame and a nonzero radial acceleration at the initial moment. 
 
Figure\,\ref{fig:4} presents the change in morphology of the allowable region during the collisional process of the green route shown in Fig.\,\ref{fig:3}. The intersection curves between the zero-velocity surface and the asteroid surface clearly demonstrate the shrinking of the allowable region where collisions are permitted. The first collision of the particle occurred within the areas beyond the zero-surface (outlined in yellow in Fig.\,\ref{fig:4}a), which was determined by the current Jacobi integral $C_0$. After the first collision, areas within allowable ranges shrank a little toward the minimum geopotential point due to the reduction of Jacobi integral (Fig.\,\ref{fig:4}b). Figures\,\ref{fig:4}c and d illustrate the shrinking of the allowable regions after the second and third collisions. As can be seen, the allowable region vanished from the eastern hemisphere after the second collision and shrank further after the third collision. We also note that with the Jacobi integral dropping, the zero-velocity surface extended outward on the equatorial plane. Thus, it is reasonable to speculate that the lowest geopotential region on the surface is usually located around the bulge of the highest elevation area at the equator.

The rest of this paper focuses on the propagation of the particle cloud shed from the asteroid surface based on the understanding of the saltation dynamics of a single particle. Figure\,\ref{fig:5} presents how the collision sites change with the increasing number of collisions for the coefficient of restitution $\varepsilon=0.45 $ and shows a color map indicating the geopotential on the asteroid surface. While the number of collision sites is equal in all four panels of Fig.\,\ref{fig:5}, the density is different. This figure illustrates the common tendency among all eight simulation groups. At the initial moment, the particles are uniformly distributed in the unstable region, which is encompassed by the contour plot of the geopotential with $V(\vec{r})=-0.1364\,\mathrm{J \, kg^{-1}}$, as shown by Fig.\,\ref{fig:5}a. After the first collision, differing from the initial distribution, the collision sites are concentrated in three areas surrounded by the geopotential contour with a value of $-0.1374\,\mathrm{J \, kg^{-1}}$, as shown by Fig.\,\ref{fig:5}b. After the second collision, the sites become further concentrated toward the middle of the three areas mentioned earlier, shown in Fig.\,\ref{fig:5}c. After several collisions, we found all particles completed the saltation process of frequent collisions with the surface and then entered the long-term cycling orbits (for the simulated time span). These orbits were all launched with low speeds from a small area with the minimum geopotential, as shown in Fig.\,\ref{fig:5}d. Hereafter, we refer to these orbits as "non-return orbits." The result illustrates that the collision events gradually consume the generalized energy of the saltation particles and drive the saltation trajectories toward a unified destination, that is, the low-geopotential region. This process agrees with the theoretical analysis we outlined. Thus, although the particles originated from different locations in the unstable region, their non-return orbits always launched from a small area where the geopotential was the lowest on the overall surface. We also note that the Jacobi integral of the non-return orbits became reduced to near the minimum value, and almost the entire surface of the asteroid fell into the forbidden region. We speculate that the trajectories will never return to the asteroid surface again unless the trajectories fall back to the same small launching area. If this case were to occur, the trajectory of the particle would be an approximate periodic motion.
  
Based on the results shown in Fig. \ref{fig:5}, we can sketch out a common evolutionary path of the particle saltation routes in the vicinity of a fast-rotating asteroid, that is, Particles dislodged from the unstable region may experience collisions with the surface that cause a declination of the particle's generalized energy and a shrink of accessible regions on the asteroid surface. After a sufficient number of collisions, all particles will enter non-return orbits that launch from the minimum geopotential area of the unstable region. There are actually three types of dynamical fates for the non-return orbits: i) a permanent cycling motion around the asteroid, ii) a return to the launching area (with nearly zero velocity), or iii) an escape from the gravitational influence of the asteroid. The first two outcomes lead to a cumulative effect of the shed mass near the asteroid and possibly a collisional growth of particles, and the third type leads to a mass leaking effect that finally cleans up the vicinity of the asteroid. Importantly, our simulation is based on the movement of the particles over a period of one month, but the three possible dynamical fates of the shed particles would result from the particles undergoing long-term evolution, so we cannot provide the statistical data for each dynamical fate. This will be the focus of our next work.
    
\subsection{Influence of the surface contact property}\label{3.2}

We examine the mechanical energy per unit mass of the particle cloud for all eight groups of sample particles, hereafter abbreviated as "average mechanical energy".. For the Keplerian motion in the inertial frame, the mechanical energy determines the orbital semi-major axis. Thus, the average mechanical energy can be taken as a qualitative indicator for the expansion of the particle cloud considered in this paper. This section focuses on the dependence of the shed particle expansion on the surface contact property. We chose a simplified factor, the coefficient of restitution, to represent the contact mechanism and investigated the temporal-spatial propagation of the particle cloud on different values of the coefficient of restitution. Figure\,\ref{fig:6} shows the evolution of the average mechanical energy of the particle cloud for the eight groups of sample particles with different coefficients of restitution. In this paper, the average mechanical energy of the particle cloud is defined as
    
 \begin{equation}\label{7}
    	\overline{ME}=\frac{1}{2\mathrm{n}}\sum_{i=1}^{\mathrm{n}}(\vec{\omega}_i\times\vec{r}_i+\dot{\vec{r}}_i)\cdot(\vec{\omega}_i\times\vec{r}_i+\dot{\vec{r}}_i)+U(\vec{r}_i) 
 ,\end{equation}
where $\mathrm{n}$ indicates the number of sample particles and $i$ represents $the$ $ith$ particle.

As demonstrated in Fig.\,\ref{fig:6}, the average mechanical energy shows an overall increase with time regardless of the value of $\varepsilon$. We note that although the average mechanical energy curves correspond to different coefficients of restitution that cover a wide range (i.e., $0.15\sim0.85$), they increase at almost the same rate. This means that the average mechanical energy is not sensitive to the coefficient of restitution. The mechanical energy differences caused by different coefficients of restitution at five different times, $0.1$, $7$, $14$, $21$ and $29$ days, are all below $4\%$ in terms of the average value. This result implies that the orbital semi-major axis distribution of the particle cloud shed from a critical rotating asteroid is potentially independent of collision property of regolith grains.

The fact that the particle cloud with different coefficients of restitution spread outward with almost identical orbital semi-major axes seems to be an anomaly because different coefficients of restitution lead to different rebound velocities, and different rebound velocities inevitably produce different orbits and orbital energy. In order to discover the intrinsic mechanism of this result, we checked the mechanical energy curves for particles with different coefficients of restitution, as shown in Fig.\,\ref{fig:7}. Figure\,\ref{fig:7}a shows the global change in the mechanical energy curve over the 30-day simulation time, and Fig.\,\ref{fig:7}b shows a locally enlarged view of the first $103,500$ seconds of Fig.\,\ref{fig:7}a.

The mechanical energy increments shown in Figs.\,\ref{fig:6} and \ref{fig:7} come from two mechanisms: first, the work of the collisional forces when the particles come in contact with the asteroid surface and, second, the work of the gravitational force due to the spinning asymmetry gravitational field. We note that the change caused by collision is an instantaneous process and the gravitational potential does not change. Thus, comparing the particle velocities before and after collisions, the change in mechanical energy caused by collision is
\begin{equation}\label{9_1}
	\Delta{ME}=(\vec{\omega}\times\vec{r})\cdot\Delta{\dot{\vec{r}}}+\frac{1}{2}(\varepsilon^2-1)\vert \dot{\vec{r}}^{-} \vert^2
,\end{equation}
where the $\Delta{\dot{\vec{r}}}= \dot{\vec{r}}^{+}-\dot{\vec{r}}^{-}$ and the $\dot{\vec{r}}^{+}$ and $\dot{\vec{r}}^{-}$ indicate the particle velocity with respect to the body-fixed frame after and before the collision, respectively. Thus, depending on the particle velocity, local surface topography, and the coefficient of restitution, the change in mechanical energy caused by the collision may increase or decrease. 

Regarding the other aspect, the asymmetric gravitational field provides a possibility for a particle to be ejected into a hyperbolic escape trajectory or, conversely, for a hyperbolic orbit to be captured into an elliptic orbit  \citep{Scheeres_1996}. Hence, the mechanical energy of the particle may change due to the interaction with the spinning asymmetry gravitational field. Specifically, the irregularly shaped asteroid induces an asymmetric gravitational field that breaks the conservation of the mechanical energy. Thus, even if the particle does not collide with the asteroid, its mechanical energy still changes while orbiting around the asteroid, and the changing rate depends on its relative position regarding the body-fixed frame of the asteroid. Combining Eqs.\,(\ref{2}) and \,(\ref{5}), the mechanical energy of a particle can be expressed as
\begin{equation}\label{12}
	ME=J+(\vec{\omega}\times\vec{r})\cdot(\vec{\omega}\times\vec{r}+\dot{\vec{r}})
.\end{equation}

Noting that $\dot{J}=0$, the time derivative of the mechanical energy yields
\begin{equation}\label{13}
\frac{\mathrm{d} ME}{\mathrm{d} t}=-(\vec{\omega}\times\vec{r})\cdot{\nabla U(\vec{r})}
.\end{equation}
The details of this derivation can be found in \cite{Yu_Resonant_2013}. Equation\,(\ref{13}) defines a scalar field in the body-fixed frame. Since $(\vec{\omega}\times\vec{r})$ and $\nabla U(\vec{r})$ are not always perpendicular in a asymmetric gravitational field, the time variation determined by Eq.\,(\ref{1}) is nonzero for most of the spatial region and is particularly rapid in the vicinity of the asteroid. 

From the view of an individual particle, its mechanical energy changes smoothly as the particle orbits around the asteroid, and it changes abruptly every now and then when the particle collides with the asteroid surface, as shown in Fig.\,\ref{fig:7}b. Although we did observe that particles can acquire energy through collision with some special terrain (such as with a surface protrusion), this situation is very rare. Most collisions reduce the mechanical energy, as shown in the red dashed box in Fig.\,\ref{fig:7}b. However, we noticed the mechanical energy loss due to collision becomes flooded in a short time by the constant change from the asymmetric gravitational field. In statistics, the total mechanical energy of all simulated particles maintains an increasing trend, as shown in Fig.\,\ref{fig:6}.

We also note that the mechanical energy drops do have a negative correlation with the coefficient of restitution within a short period of time after the first collision, as shown in the blue dashed box in Fig.\,\ref{fig:7}b. However, as the orbits of the sampled particles continued to evolve, this short-term correlation became rapidly flooded because the particle orbits close to the asteroid exhibited unstable and even chaotic behavior that resulted in the frequent change of orbital energy \citep{Scheeres_Dynamics_Ellipsoids_1994}. After sufficient time, the behaviors of the mechanical energy curves became adequately diffused, a process which is governed by the chaotic nature of the orbital dynamics near asteroids rather than by collisions. Thus, over a relatively long term, no monotonic correlation between the value of the mechanical energy and the coefficient of restitution was identified, as shown in Fig.\,\ref{fig:7}a. This gives a basic reason why the average mechanical energy of the particle cloud exhibits no statistical dependency on the coefficient of restitution and shows convergence despite the coefficients of restitution being different.
      
Meanwhile, we also noted that as the coefficient of restitution decreases, the loss in the generalized energy caused by a single collision becomes larger, according to Eq.\,(\ref{8}). Statistically, this implies that fewer collisions are needed to translate the particle to the minimum geopotential region when the coefficient of restitution is smaller. As a validation, Fig.\,\ref{fig:8} counts the total number of collisions for all eight groups of sample particles, showing a conclusion consistent with the tendency.

The analysis of the saltation routes shows that regardless of the value of the coefficient of restitution, the saltation routes always converge toward the vicinity of the region of minimum geopotential within the unstable region and then from that region into the non-return orbits. Therefore, based on whether the particle enters the non-return orbits, the orbit of the particle can be divided into the saltation orbits and the non-return orbits. According to the conclusions in Sec. \ref{3.1}, there are two ways for particles to enter the non-return orbits: One is to enter a non-return orbit directly from their initial location near the minimum geopotential. The second is to collide with the surface several times and then enter a non-return orbit due to the generalized energy gradually dropping to around the minimum value. Therefore, all the saltation particles eventually transfer to non-return orbits through a common dynamical route, and the mass transfer ratio (particle mass transferred to the non-return orbit per unit time) determines the density distribution of the particle cloud in the vicinity of the asteroid.

Statistics were calculated based on the results of simulations ran with the eight groups of sample particles. These statistics showed the mass transfer ratio of the shed particle cloud as a function of the coefficient of restitution. Figure\,\ref{fig:9} counts the time points at which the sampled particles enter non-return orbits. The broken lines in Fig.\,\ref{fig:9} show that all the coefficients of restitution lead to a similar trend, that is, over 90\% of the shed particles are transferred to non-return orbits within 15 days, and thus the simulated time of 30 days is sufficiently long to reflect the full life cycles of the particles. Histograms in Fig.\,\ref{fig:9} show that the frequency peak of a particle entering a non-return orbit appears on the first day for all simulations, and peak values decrease with the increase in coefficients of restitution. This indicates that the flux of the particle transferred to a non-return orbit peaks at the beginning of the shedding process and drops rapidly with time, implying that a dense particle cloud may emerge early during the shedding process and subsequent collisional growth of particles may also occur in this stage (depending on the spatial density and the contact properties of the particles). This result that peak values decrease with the increase in coefficients of restitution also illustrates that smaller coefficients of restitution lead to a greater flux of particles transferred to non-return orbits.

\section{Discussion and conclusion} \label{S_4}
This paper explored the complex saltation processes of loose particles on the surface of a fast-rotating asteroid, with a special focus on the parameter dependency of such processes. Eight groups of simulations with different coefficients of restitution were performed with a sufficiently large sample size, which demonstrates the influence of contact properties (the coefficient of restitution) on the dynamical propagation of a shed particle.

Although we consider a wide parameter range for the coefficients of restitution, we find a general evolutionary trend for the saltation particles shed from the surface of fast-rotating asteroids, namely, particles dislodged from the unstable region will enter non-return orbits that are connected to the minimum geopotential area of the unstable region after a sufficient number of collisions. This tendency remains constant for a large group of asteroids near the critical spin limits. We note that due to the irregular shape of the asteroid, the geopotential energy of the unstable region may be not uniformly smaller than that of the stable region, that is, a small part of the stable region will possibly exist in the allowable region determined by the effective potential energy value of the unstable region. Thus, the saltation orbits do not technically have to end in the non-return orbits. Theoretically, there could be particles that are transferred to the stable region and become stuck to the surface. However, in our simulations, all the particles eventually entered non-return orbits, meaning the ratio of shed particles re-accreted onto the asteroid surface is statistically very small.
   
Furthermore, analysis of the zero-velocity surfaces show the dynamical fates of the shed particles are closely correlated with the asteroid's surface morphology, the asymmetry gravitational field, and the rotation state, which are three factors that dominate the location of the minimum geopotential area. For rapidly rotating top-shaped asteroids, such as the primary of Didymos in this paper, the minimum geopotential usually occurs near the point of the high elevation on its equatorial ridge. This result suggests that if we determine the shape of the asteroid, we can make a qualitative assessment of the migration path of the material on or near its surface, which further allows us to predict the debris circumstance generated by shedding events occurring on such asteroids. This finding might provide support for ongoing or future asteroid missions. For example, a cumulative analysis based on the common trend of the migration paths will give benchmarking results on the formation of a particle cloud and estimates on the potential risk to spacecrafts.
   
As for the planetary system formation, assessing how the shedding events determine the subsequent temporal-spatial distribution of the particle cloud around the asteroid is of particular interest. Our simulations indicate that the orbits of the particle cloud continue to spread outward under the combined effect of the centrifugal and gravitational fields, and their orbital semi-major axis distribution is not statistically sensitive to the coefficient of restitution. However, the transfer rate of the particle from the saltation orbits to the non-return orbits is affected by the coefficient of restitution, according to the result from Fig.\,\ref{fig:9}. Specifically, a smaller coefficient of restitution corresponds to faster transferring, which results in a relatively denser particle cloud. However, a smaller coefficient of restitution also leads to greater energy dissipation in inter-particle collisions, which is conducive to the aggregation of particles. Therefore, we speculate that a smaller coefficient of restitution is more favorable to the collisional growth of particle aggregates, which are the original building blocks for asteroid satellites. Nevertheless, the role that the coefficient of restitution plays in the generation of asteroid moonlets could be more complex. A smaller coefficient of restitution contributes to the accumulation of particles in multiple ways, as it not only causes larger collisional dissipation but also changes the temporal-spatial distribution of the particle cloud.
   
The results of this study sketch an evolutionary path of the saltation particle and represent a common trend for regolith particles shed from totally different locations. We take this trend as an invariant property for a wide variety of asteroids near their critical spin limits, and we speculate that the dynamical fate of the described path actually determines the spread-accumulation behavior of the particle cloud, which provides a new idea for further investigations into why some top-shaped asteroids have a moon while others do not. Specifically, since the regolith particles shed from the spin-unstable asteroids have the same migration laws, we can then focus our future research on the material-loss process caused by this trend. However, as to whether this process can lead to the aggregation and growth of satellites, this article cannot provide an answer. Theoretically, losing materials through the escape path is easier than through the other two paths, so the escape path may be more detrimental to the formation of satellites. However, the specific spread-accumulation process cannot be explained completely by qualitative analysis, and the spread process needs to be combined with the quantitative calculation of the discrete element method to give some parameter intervals that satellites can be formed. This will also be the focus of our next work. Furthermore, further investigation will also be devoted to exploring the topographic dependence of the shed particle cloud evolution, which is expected to add new details to the formation theory of binary-triple asteroid systems.
   
    \begin{acknowledgements}
    	We thank the members of Hera WG3 group for the constructive conversations. Y.Y. acknowledges financial support provided by the National Natural Science Foundation of China Grant No. 12022212. 
    \end{acknowledgements}
    
    \bibliographystyle{aa}
    \bibliography{manuscript}
\end{document}